\RequirePackage{fix-cm}
\documentclass[twocolumn,epjc3]{svjour3}  
\smartqed  
\RequirePackage[dvipdfmx]{graphicx}
\RequirePackage{latexsym}
\RequirePackage[numbers,sort&compress]{natbib}
\RequirePackage[colorlinks,citecolor=blue,urlcolor=blue,linkcolor=blue]{hyperref}
\usepackage{array}
\usepackage{braket}
\usepackage{bm}
\usepackage{subfigure}
\usepackage{verbatim}
\usepackage{wrapfig}
\usepackage{ascmac}
\usepackage{makeidx}
\usepackage{ulem}
\usepackage[version=3]{mhchem}
\usepackage{here}
\usepackage{comment}
\newcommand{\JSNS}{JSNS$^2$\,}

\newcommand{\Sterile}{$\bar{\nu}_{\mu} \to \bar{\nu}_{e}$~}

\newcommand{\micro}{$\mu \mathrm{s}$\,}

\newcommand{\eVsq}{$\mathrm{eV}^2$}

\journalname{Eur. Phys. J. C}
\begin{document}

\title{
Study on the accidental background of the \JSNS experiment
}

\author{
D.~H.~Lee\thanksref{e1, addr1} \and
S.~Ajimura\thanksref{addr2} \and
M.~K.~Cheoun\thanksref{addr3} \and
J.~H.~Choi\thanksref{addr4} \and
J.~Y.~Choi\thanksref{addr5} \and
T.~Dodo\thanksref{addr6, addr9} \and
J.~Goh\thanksref{addr7} \and
K.~Haga\thanksref{addr8} \and
M.~Harada\thanksref{addr8} \and
S.~Hasegawa\thanksref{addr9, addr8} \and
T.~Hiraiwa\thanksref{addr2} \and
W.~Hwang\thanksref{addr7} \and
H.~I.~Jang\thanksref{addr5} \and
J.~S.~Jang\thanksref{addr10} \and
H.~Jeon\thanksref{addr11} \and
S.~Jeon\thanksref{addr11} \and
K.~K.~Joo\thanksref{addr12} \and
D.~E.~Jung\thanksref{addr11} \and
S.~K.~Kang\thanksref{addr13} \and
Y.~Kasugai\thanksref{addr8} \and
T.~Kawasaki\thanksref{addr14} \and
E.~J.~Kim\thanksref{addr15} \and
J.~Y.~Kim\thanksref{addr12} \and
S.~B.~Kim\thanksref{addr16} \and
W.~Kim\thanksref{addr17} \and
H.~Kinoshita\thanksref{addr8} \and
T.~Konno\thanksref{addr14} \and
I.~T.~Lim\thanksref{addr12} \and
C.~Little\thanksref{addr18} \and
E.~Marzec\thanksref{addr18} \and
T.~Maruyama\thanksref{addr1} \and
S.~Masuda\thanksref{addr8} \and
S.~Meigo\thanksref{addr8} \and
S.~Monjushiro\thanksref{addr1} \and
D.~H.~Moon\thanksref{addr12} \and
T.~Nakano\thanksref{addr2} \and
M.~Niiyama\thanksref{addr19} \and
K.~Nishikawa\thanksref{addr1} \and
M.~Y.~Pac\thanksref{addr4} \and
H.~W.~Park\thanksref{addr12} \and
J.~S.~Park\thanksref{e2, addr17} \and
R.~G.~Park\thanksref{addr12} \and
S.~J.~M.~Peeters\thanksref{addr20} \and
C.~Rott\thanksref{addr21, addr11} \and
K.~Sakai\thanksref{addr8} \and
S.~Sakamoto\thanksref{addr8} \and
T.~Shima\thanksref{addr2} \and
C.~D.~Shin\thanksref{addr1} \and
J.~Spitz\thanksref{addr18} \and
F.~Suekane\thanksref{addr6} \and
Y.~Sugaya\thanksref{addr2} \and
K.~Suzuya\thanksref{addr8} \and
M.~Taira\thanksref{addr1} \and
Y.~Yamaguchi\thanksref{addr8} \and
M.~Yeh\thanksref{addr22} \and
I.~S.~Yeo\thanksref{addr4} \and 
C.~Yoo\thanksref{addr7} \and 
I.~Yu\thanksref{addr11}
}

\thankstext{e1}{e-mail: leedh@post.kek.jp}
\thankstext{e2}{e-mail: jungsicpark@knu.ac.kr}


\institute{%
High Energy Accelerator Research Organization (KEK), Tsukuba, Ibaraki, JAPAN \label{addr1} \and
Research Center for Nuclear Physics, Osaka University, Osaka, JAPAN \label{addr2} \and
Department of Physics, Soongsil University, Seoul 06978, KOREA \label{addr3} \and
Laboratory for High Energy Physics, Dongshin University, Chonnam 58245, KOREA \label{addr4} \and
Department of Fire Safety, Seoyeong University, Gwangju 61268, KOREA \label{addr5} \and
Research Center for Neutrino Science, Tohoku University, Sendai, Miyagi,
JAPAN \label{addr6} \and
Department of Physics, Kyung Hee University, Seoul 02447, KOREA \label{addr7} \and
J-PARC Center, JAEA, Tokai, Ibaraki JAPAN \label{addr8} \and
Advanced Science Research Center, JAEA, Ibaraki JAPAN \label{addr9} \and
Gwangju Institute of Science and Technology, Gwangju, 61005, KOREA \label{addr10} \and
Department of Physics, Sungkyunkwan University, Gyeong Gi-do, KOREA \label{addr11} \and
Department of Physics, Chonnam National University, Gwangju, 61186, KOREA \label{addr12} \and
School of Liberal Arts, Seoul National University of Science and Technology, Seoul, 139-743, KOREA \label{addr13} \and
Department of Physics, Kitasato University, Sagamihara 252-0373, Kanagawa, JAPAN \label{addr14} \and
Division of Science Education, Jeonbuk National University, Jeonju, 54896, KOREA \label{addr15} \and
School of Physics, Sun Yat-sen (Zhongshan) University, Guangzhou, 510275, China
\label{addr16} \and
Department of Physics, Kyungpook National University, Daegu 41566, KOREA \label{addr17} \and
University of Michigan, Ann Arbor, MI, 48109, USA \label{addr18} \and
Department of Physics, Kyoto Sangyo University, Kyoto, JAPAN \label{addr19} \and
Department of Physics and Astronomy, University of Sussex, Brighton, UK \label{addr20} \and
Department of Physics \& Astronomy, The University of Utah, UT, 84112, USA \label{addr21} \and
Brookhaven National Laboratory, Upton, NY, 11973-5000, USA \label{addr22}
}

\date{Received: date / Accepted: date}

\maketitle

\begin{abstract}
\JSNS (J-PARC Sterile Neutrino Search at J-PARC Spallation Neutron Source) is an experiment that searches for sterile neutrinos via the observation of \Sterile appearance oscillations using muon decay-at-rest neutrinos.
The \JSNS experiment performed data taking from 2021.
In this manuscript, a study of the accidental background is presented. The rate of the 
accidental background is (9.29$\pm 0.39) \times 10^{-8}$ / spill with 0.75 MW beam power 
and comparable to the expected number of signal events. 
\end{abstract}

\section{Introduction} 
\label{intro}
The existence of sterile neutrinos has been a crucial issue in the neutrino physics community for over 20 years. The experimental results from~\cite{cite:LSND, cite:GALLEX, cite:SAGE, cite:RA, cite:MiniBooNE2013, cite:MiniBooNE2018} could be interpreted as indications of the existence of sterile neutrinos with a mass-square differences of around 1~\eVsq. 

The \JSNS experiment, proposed in 2013~\cite{cite:proposal}, is designed to search for neutrino oscillations due to a sterile neutrino at the Material and Life science experimental Facility (MLF) in J-PARC.
MLF provides an intense and high-quality neutrino flux of 
$1.8 \times 10^{14}\,\nu$/year/cm$^2$, from muon decay-at-rest ($\mu$DAR)
produced using a 1~MW proton beam with a 25~Hz repetition rate~\cite{cite:TDR}.
The neutrinos are produced by injecting 3~GeV protons from  a rapid cycling synchrotron onto a mercury target in the MLF.
The experiment uses a Gadolinium (Gd) loaded liquid scintillator (Gd-LS) detector with 
0.1~w\% Gd concentration placed at 24~m from the target.

The \JSNS experiment aims to perform a direct test of the 
LSND observation~\cite{cite:LSND}.
The same experimental principle used by the LSND experiment~\cite{cite:LSND} to try and observe the \Sterile oscillation is used: inverse beta decay (IBD). 
There are several improvements offered by the \JSNS experiment.
In order to identify IBD events, a delayed coincidence between the positron signal (prompt signal: up to 53~MeV) and neutron capture signal is used for selection. Gd is used to identify neutron captures. After capturing thermal neutrons on Gd, neutron capture on Gd generates gamma-rays with higher energies and shorter capture times ($\sim$8~MeV, $\sim$30~$\mu$s) than neutron capture on hydrogen (2.2~MeV, $\sim$200~$\mu$s).
Therefore, accidental backgrounds coincident in the delayed signal region can be reduced by $\sim 6$ times compared to the hydrogen capture used in the LSND experiment, due to the shorter capture time.
In addition, the short-pulsed beam, two 100~ns pulses in a 600~ns interval in each spill with a repetition of 25~Hz, enables us to set a timing window for the IBD prompt signal to 2.0 to 10~\micro\ from the proton beam collision so that the neutrinos from pion and kaon decay and fast neutrons generated at the target can be rejected efficiently.
However, the efficiency for the $\mu$DAR neutrinos can be kept at 48\% because of the muon lifetime ($2.2~\mathrm{\mu s}$).
The cosmogenic background is also reduced by a factor of $10^{-4}$.

For \JSNS, understanding the accidental background is essential since it is  one of main 
backgrounds for the sterile neutrino search. Another important background, the correlated background, is 
described in \cite{cite:EPJC_Hino}.      
A detailed discussion about the signal detection principle and the background rejection technique 
can be found in~\cite{cite:proposal, cite:TDR}.

\section{Setup}
\label{data_taking}
The \JSNS experiment has been taking data with a single detector since 2021. 
Except for the beam maintenance period, which typically corresponds to a few months over the summer, the data has been accumulated continuously. 
The proton beam power has been increased from 600~kW in 2021 to 840~kW in 2023. 
There is usually a one-day facility maintenance per week and we continue to take data during that time to acquire beam-off data.
The integrated number of proton-on-target (POT) collected was $2.94 \times 10^{22}$, 
corresponding to less than $\sim$28.0\% of the required POT of the \JSNS experiment. 

To understand the rate and properties of the accidental background, dedicated calibration 
data were taken.
The accelerator-driven timing signal was used for the trigger~\cite{cite:daq, cite:detector}, which caused a 
125~$\mu$s time window for data acquisition system (FADC) to be opened, with no energy bias. 
The total available number of spills is 2.46 $\times 10^{6}$, therefore 
the expected number of IBD signal events from this calibration run is estimated 
to be much less than 1~\cite{cite:TDR}. Thus, all observed events are likely to be background.

\subsection{Experimental setup}
\label{setup}
The \JSNS detector is a cylindrical liquid scintillator detector with 4.6~m diameter and 3.5~m height located at a distance of 24~m from the mercury target of the MLF. It consists of 17~tonnes of Gd-LS contained in an acrylic vessel, and 33~tonnes unloaded liquid scintillator (LS) in a layer between the acrylic vessel and a stainless steel tank.
The LS and the Gd-LS consist of LAB (linear alkyl benzene) as the base solvent, 3 g/L PPO (2,5-diphenyloxazole) as the fluor, and 15 mg/L bis-MSB (1,4-bis(2-methylstyryl) benzene) as the wavelength shifter.
The LS volume is separated into two independent volumes by an optical separator.
The region inside the optical separator, called the ``inner detector", consists of the entire 
volume of the Gd-LS and $\sim$25~cm thick LS layer.
Scintillation light from the inner detector is observed by 96 Hamamatsu R7081 photomultiplier tubes (PMTs) each with a 10-inch diameter.
The outer layer, called the ``veto layer", is used to detect cosmic-ray induced particles coming into the detector.
A total of 24 of 10-inch PMTs are set in the veto layer. On the whole inner surfaces of the veto layer, reflection sheets are attached in order to improve the collection efficiency of the scintillation light.

\subsection{Data acquisition and trigger system}
\label{daq}
PMT signal waveforms from both the inner detector and the veto layer are digitized and recorded at a 500~MHz sampling rate by 8-bit flash analog-to-digital converters (FADCs).
As a trigger, we utilize a 25 Hz periodic signal from the accelerator scheduled timing which directs 
the proton beam towards the MLF target, called the ``kicker trigger".
The width of the acquired waveform in this trigger scheme is set to 125~\micro, which fully covers the prompt and delayed signal timing window of the IBD events.
The rate limitation of the \JSNS data acquisition system under this condition is 5~Hz, 
thus, a pre-scaling factor of 5 is required.
We mainly used this trigger and data acquisition to obtain the beam-on data for an accidental background estimation within the sterile neutrino search.

Detailed description of the detector and the triggers are given in~\cite{cite:detector, cite:daq}, respectively.

\section{Event selection}
\label{analysis}
The data set used for the background estimation was obtained using the kicker trigger.
The total number of beam spills was 2460399, equivalent to $\sim$\textcolor{black}{6}~days of data taking
with a pre-scale factor of five.
Given the readout 125~\micro time, the obtained waveforms contain multiple events.
We therefore used an event definition based on the number of hit PMTs in order to 
extract each event.
We constructed a hit time and charge series at each trigger by accumulating hit information 
along the FADC window with 60~ns coincidence width over all the PMTs.
The coincidence width was determined by considering a typical PMT pulse shape and a 
safety factor from timing calibration.
The event discrimination threshold is set to 10 hits and 50 p.e., which corresponds 
to an energy well below 1~MeV.

The event vertex position and energy reconstruction is performed 
simultaneously based on a maximum-likelihood algorithm using the charge response of each PMT.
Both vertex position and energy were calibrated by deploying a $^{252}$Cf source and 
using the 8~MeV peak in the energy spectrum resulting from thermal neutron capture on Gd (nGd).
The reconstruction performance can be found in~\cite{cite:PAC31} for nGd events and in~\cite{cite:HJKPS2020} for events with up to 60~MeV using Michel electrons.

The selection criteria and estimated efficiency are given in Table~\ref{tab:selec_eff} 
and a more detailed description is found in~\cite{cite:TDR}. 
Each efficiency in Table~\ref{tab:selec_eff} is for each corresponding criterium only.

The single rates of the prompt and the delayed candidates are separately estimated using 
the energy and the timing selections shown in Table~\ref{tab:selec_eff}. 
Note that a time difference from the beam collision timing, 
$5\le \Delta t_{\mathrm{beam-d}}\le 105$~\micro\ is used for  
the delayed single rate calculation 
instead of the selection in Table~\ref{tab:selec_eff}.
These single rates are multiplied to predict the number of the accidental
backgrounds. 
The fiducial volume is defined with $R < 140$ cm and $|z| <$ 100 cm region to avoid external backgrounds. Note that origin of the coordinate system is the center of the detector, 
and $R$ is defined as $R = \sqrt{x^2 + y^2}$.

\begin{table}[t]
    \centering
    \caption{The IBD selection criteria and their efficiencies in the \JSNS experiment~\cite{cite:TDR}. 
    The single rates of the prompt and the delayed candidates are estimated separately using 
    the energy and the timing selection in this table.    
    Note that the timing window: $5\le \Delta t_{\mathrm{beam-d}}\le 105$~\micro is used for  
    the delayed single rate calculation. 
    Each efficiency is for to each corresponding criterium only.
    \label{tab:selec_eff}}
    \vspace{3pt}
    \begin{tabular}{cc}\hline
        Requirement & Efficiency /\% \\\hline
        --Prompt Event--\\

        $20\le E_{\mathrm{p}}\le 60$~MeV                    & 92 \\
        $2.0\le \Delta t_{\mathrm{beam-p}}\le 10$~\micro    & 48 \\
        Pulse Shape Discrimination                          & 99 \\
                                                            &     \\

        --Delayed Event-- \\
        $7\le E_{\mathrm{d}}\le 12$~MeV                     & 71 \\
        $\Delta \mathrm{VTX}_{\mathrm{OB-d}}\ge 110$~cm     & 98 \\
                                                            &     \\

        --IBD paired Event-- \\
        $\Delta t_{\mathrm{p-d}}\le 100$~\micro             & 93 \\
        $\Delta \mathrm{VTX}_{\mathrm{p-d}}\le 60$~cm       & 96 \\
        Timing likelihood                                   & 91 \\\hline
        Total                                               & 25 \\\hline
    \end{tabular}
\end{table}

The prompt IBD candidates are selected using a time difference 
from beam collision timing ($\Delta t_{\mathrm{beam-p}}$) and its energy ($E_{\mathrm{p}}$).
We applied the following requirements; $2\le \Delta t_{\mathrm{beam-p}}\le 10$ \micro~and $20\le E_{\mathrm{p}}\le 60 $ MeV, in order to fully cover $\mu$DAR neutrinos from the mercury target.
The timing selection rejects beam-induced fast neutrons in the beam on-bunch timing ($0 \le \Delta t_{\mathrm{beam-p}}\le 1.5~\mathrm{\mu s}$) as well as neutrino backgrounds from kaon and pion decay whose lifetimes are 12~ns and 26~ns, respectively.

In order to identify neutron captures on Gd, the 8~MeV peak of the delayed signal energy ($E_{\mathrm{d}}$) is selected using a requirement of $7\le E_{\mathrm{d}}\le 12 $ MeV~\cite{cite:SRBG}.
There are nGd events associated with fast neutrons induced by the beam contributing to the IBD delayed candidates as an accidental background~\cite{cite:BG2014}.
Pulse Shape Discrimination (PSD) is used within the IBD prompt timing region to reduce neutrons induced by cosmic rays as shown in~\cite{cite:EPJC_Hino}. However, it is optimized for the energy range of $20\le E_{\mathrm{p}}\le 60$~MeV. Thus, we need another strategy to reject those nGd events.
Since these nGd events correlate spatially with an activity made by beam-induced fast neutrons, we can reject them with the spatial correlation.
In particular, we applied a requirement on the spatial difference between on-bunch events and delayed candidates, $\Delta \mathrm{VTX}_{\mathrm{OB-d}}\ge 110$ cm.
The distribution of $\Delta \mathrm{VTX}_{\mathrm{OB-d}}$ and the efficiency estimation can be found in~\cite{cite:SRBG}.
The on-bunch event tagging condition is set to $0 \le \Delta t_{\mathrm{beam-OB}} < 1.5~\mathrm{\mu s}$ and $1 \le E_{\mathrm{OB}}\le 200$ MeV.

\begin{figure}
    \includegraphics[width=0.48\textwidth]{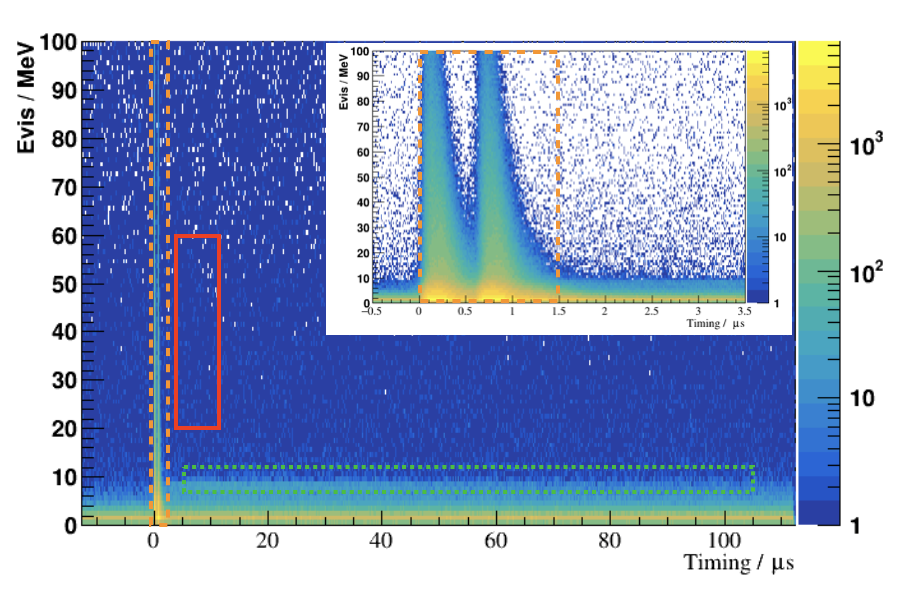}%
    \caption{A two-dimensional distribution of energy and timing before
    event selection, used to demonstrate the IBD selection region. The selected regions for the IBD prompt candidate from positron (red solid box), the IBD delayed candidate from gamma-rays resulted in thermal neutron capture on Gd (green dashed box) and the beam on-bunch event (orange dashed box) are overlaid. 
    Note that the events are shown around the prompt signal timing region. There are two event clusters within 0 to 1.5 \micro, which reflect the proton beam structure of the MLF. They are caused by neutrons produced at the mercury target.
    One can see that the IBD prompt signal region is well separated from the on-bunch region.
    The events in the IBD delayed signal region must also satisfy $\Delta t_{\mathrm{p-d}} > 0$ in the actual delayed coincidence. 
    \label{fig:e_t_def}}
\end{figure}

Figure~\ref{fig:e_t_def} demonstrates the energy and timing selection windows in a two-dimensional distribution of energy and timing. The red and green boxes represent the prompt and delayed signal 
regions. The beam on-bunch event are also defined as an orange dashed box.

The events induced from muons or incoming particles from outside are rejected from both prompt and 
delayed candidates by the veto region of the detector. 
The events which have more than 30 photo-electrons (p.e.) of a total charge  for the  
top-side 12 PMTs or 40 p.e. for the bottom-side 12 PMTs are rejected. 

The decayed Michel electrons (ME) from muon decay are also crucial background. The IBD
candidates are rejected if parent muons are found during 10 $\mu$s before the candidate. 
The events that have more than 100 p.e. for the top 12 or bottom 12 PMTs are categorized as 
parent muons except for the beam timing.
At the beam timing, incoming neutrons induced by the beam ($0.7464 \pm 0.0006$ / spill) could pass the criteria for selecting parent muons; thus we applied a different condition only during the beam timing.
The equation $E + Q_{veto}/9 > 200$ is used to define the parent muons of Michel electrons during the beam timing, where $E$ is the deposited energy in the target and catcher regions and $Q_{veto}$ is the total charge of veto region with p.e. unit.  
``$Q_{veto}/9$" converts p.e. to MeV units in the equation. For example, $Q_{veto}$ = 9 p.e. corresponds to 1 MeV in the veto region.

\begin{figure}
    \includegraphics[width=0.45\textwidth]{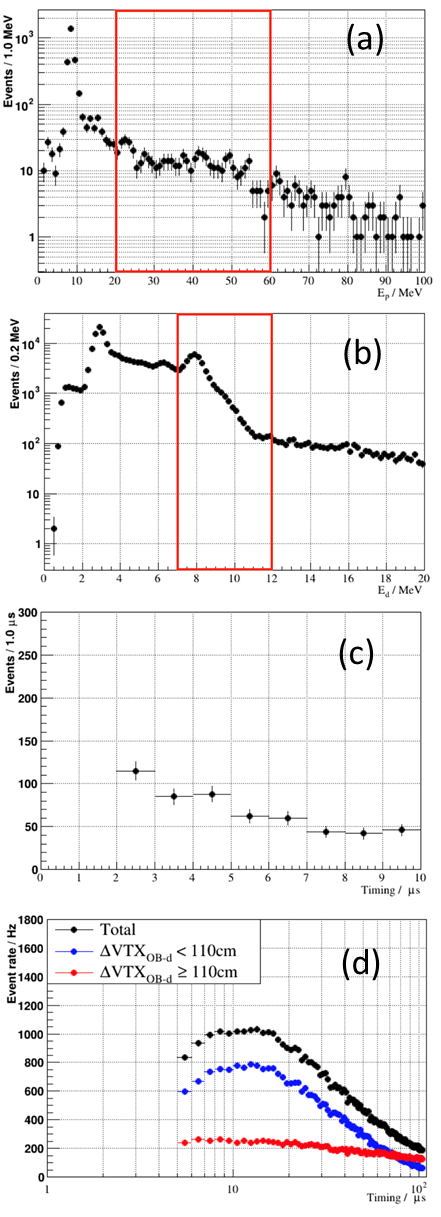}%
    \caption{The distributions of the variables used in the IBD selection: (a) $E_{\mathrm{p}}$, (b) $E_{\mathrm{d}}$, (c) $T_{\mathrm{p}}$ from beam timing and (d) $T_{\mathrm{d}}$ from beam. Each distribution has all of the selection criteria applied, except for the selection on the variable displayed. The red boxes show the selection criteria for $E_{\mathrm{p}}$ and $E_{\mathrm{d}}$.
    \label{fig:result_ibd}}
\end{figure}

Fig.~\ref{fig:result_ibd} shows the distributions of $E_{\mathrm{p}}$ (a), $E_{\mathrm{d}}$ (b) and $T_{\mathrm{p}}$ 
from beam (c), $T_{\mathrm{d}}$ from beam (d) after applying the IBD selection.
 
 The number of selected events of the IBD prompt and the delayed candidates are
542 and 44336, respectively, thus
the single rates for those are
(2.20$\pm$0.09) $\times 10^{-4}$ / spill (prompt) and (1.80$\pm$0.01) $\times 10^{-2}$ / spill (delayed), respectively.
 Note that the uncertainties mentioned in this paper contain both the statistical and systematic 
uncertainties. For the systematic uncertainties, the energy scale, the FADC timing and the reproducibility are considered. 
Figure~\ref{fig:result_ibd} (d) shows that the rejection using spatial correlation between beam neutrons 
and delayed candidates ($\Delta \mathrm{VTX}_{\mathrm{OB-d}}$) works well.  

\section{Beam on-off comparison of single rates}
\label{beam_de@}
The comparison of single rates between beam-on and beam-off are also performed.
The beam-off data is also taken using the the kicker trigger with 125 $\mu s$, and the IBD
event selection is identical to that in beam-on data. The total number of the beam-off triggers 
is 466,348, which is equivalent to about 1 day of data taking.
Table~\ref{Tab:beamdep} shows the comparison of the IBD prompt and delayed candidates.

\begin{table}[t]
    \centering
    \caption{The beam power dependence of the rate of the IBD prompt and delayed candidates.} 
    \vspace{3pt}
    \begin{tabular}{ccc}\hline
        Beam power & prompt Rate / spill & Delayed rate / spill \\\hline
          0  (beam off)  & (1.85$\pm$0.20)$\times 10^{-4}$ & (3.98$\pm$0.09)$\times 10^{-3}$ \\
          750 kW         & (2.20$\pm$0.09)$\times 10^{-4}$ & (1.80$\pm$0.01)$\times 10^{-2}$ \\\hline
    \end{tabular}
    \label{Tab:beamdep}
\end{table}

This result shows the rate of the IBD prompt background are independent from 
the beam power, which indicates that the cosmic ray induced particles are 
dominated in this region. 
On the other hand, that of the delayed background has large dependence of the beam.

\section{Accidental background rates}
The accidental background rate is evaluated by multiplying the single rates as follows:
\begin{equation}
  R_{acc.} \sim R_{p} \times R_{d} \times \epsilon_{cut} 
\label{eq:acc_calc}
\end{equation}
where $R_{acc}$ is the accidental background rate (/spill), $R_{p}$ is the single rate of 
the IBD prompt candidates (/spill), and $R_{d}$ is the rate of of delayed candidates (/spill). 
If the additional selections are applied, the selection efficiency to the accidental background 
($\epsilon_{cut}$) should be also multiplied.
As shown in Table~\ref{tab:selec_eff}, the spatial correlation selection
between the prompt and delayed candidates, 
$\Delta \mathrm{VTX}_{\mathrm{p-d}}$, and timing likelihood should also be considered. 
In this manuscript, only the spatial correlation cut is described. The likelihood
will be considered in a different future paper.

\subsection{Evaluation of the spatial cut efficiency using a spill shift method}
\begin{figure}
    \includegraphics[width=0.5\textwidth]{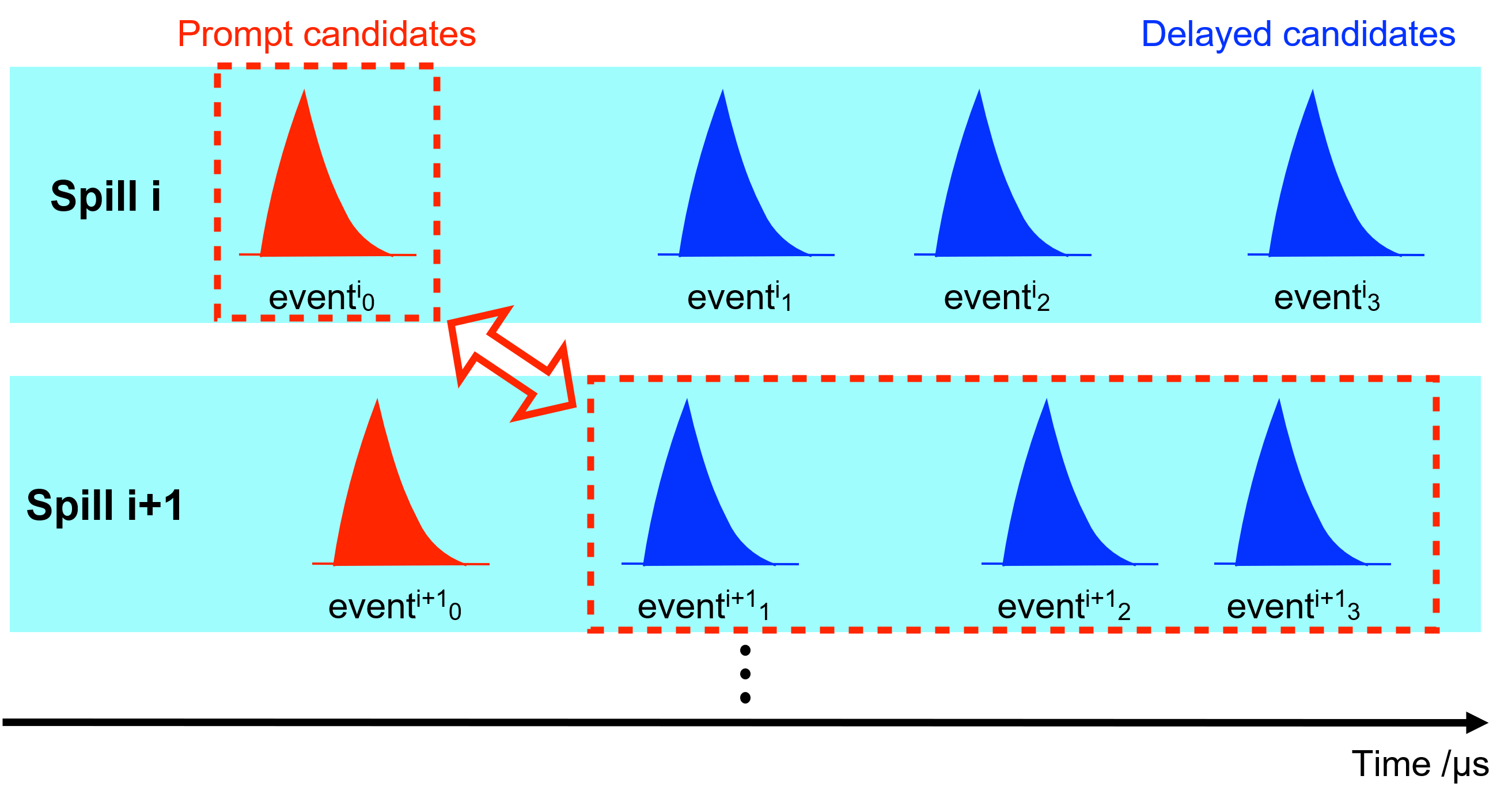}%
    \caption{
       The principle of the spill shift.  
    \label{fig:spill_shift_cartoon}} 
\end{figure}

In order to evaluate the efficiency of $\Delta \mathrm{VTX}_{\mathrm{p-d}}$ of 
the pure accidental
uncorrelated background, a novel technique is invented, which we call
``spill shift method".
Within the same beam spill, the correlated background is dominating, as discussed in 
\cite{cite:EPJC_Hino}.  However, if we use subsequent spills, beam correlated events disappear after more than 40~ms and thus a pure sample of uncorrelated accidental events is obtained. Once the IBD prompt candidate in one spill is found, 
the different beam spills are analyzed to find the paired delayed candidates.
At first, the next beam spill of the spill that have the IBD prompt candidate is used. Secondly, 
the beam spill that is different by two spills from the spill with IBD prompt
candidate is used. This process is repeated 
from the 1\textsuperscript{st} to the 10,000\textsuperscript{th} next spills to get higher statistics.
Figure~\ref{fig:spill_shift_cartoon} shows the cartoon illustrating the principle
of the spill shift in the case of 1 spill shift.

\begin{figure}
    \includegraphics[width=0.5\textwidth]{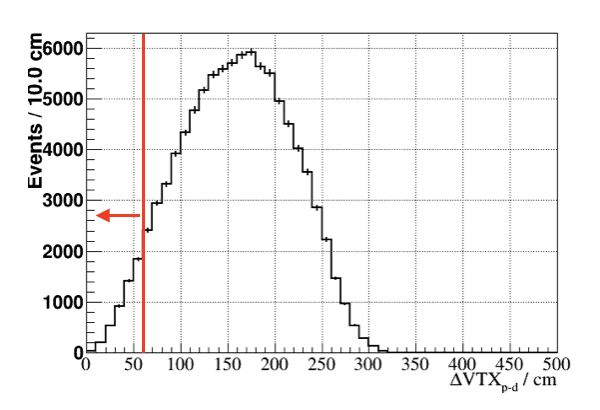}%
    \caption{
    The estimated $\Delta \mathrm{VTX}_{\mathrm{p-d}}$ distribution of the accidental 
    background with the spill shift {\color{black}method.} 
    The red line shows a cut value, and the remained fraction of the accidental background is 5.1$\pm 0.1 \%$.
    \label{fig:DVTX_acc}} 
\end{figure}

Figure~\ref{fig:DVTX_acc} shows the estimated $\Delta \mathrm{VTX}_{\mathrm{p-d}}$ distribution.
The estimated efficiency of the $\Delta \mathrm{VTX}_{\mathrm{p-d}} \le$ 
60 cm cut is 5.1$\pm 0.1 \%$.

\begin{table}[t]
    \centering
    \caption{The numbers used in Equation~\ref{eq:acc_calc} and 
    the calculated accidental background.}
    \label{tab:acci_cal}
    \vspace{3pt}
    \begin{tabular}{cc}\hline
         & Value  \\\hline
        Prompt Rate / spill & (2.20$\pm$0.09)$\times 10^{-4}$ \\
        Delayed Rate / spill & (1.80$\pm$0.01)$\times 10^{-2}$ \\
        Efficiency of $\Delta \mathrm{VTX}_{\mathrm{p-d}}\le 60$~cm       & 5.1$\pm 0.1 \%$ \\
        Efficiency of Timing likelihood ~\cite{cite:TDR}                                  & 46$\%$ \\\hline
        Accidental Rate / spill / 0.75MW                                               & (9.29$\pm0.42) \times 10^{-8}$ \\\hline
    \end{tabular}
\end{table}

\subsection{Accidental background rates}
The expected accidental background rate is (9.29$\pm 0.42) \times 10^{-8}$ / spill / 0.75MW using 
the Equation~\ref{eq:acc_calc}. The numbers used in this calculation of the accidental background rate are summarized in Table~\ref{tab:acci_cal}. 
The single rates, the spatial correlation cut efficiency 
mentioned above 
and the assumed efficiency of timing likelihood ($\sim$46\%) described in ~\cite{cite:TDR} are used.

\begin{figure}[htb!]
    \includegraphics[width=0.5\textwidth]{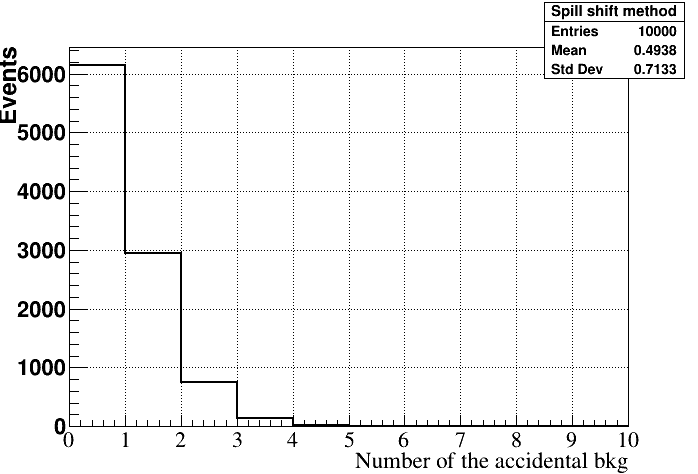}%
    \caption{
    The expected number of accidental background events from the spill shift method 
    during this calibration run with 2,460,399 spills after the 
    $\Delta \mathrm{VTX}_{\mathrm{p-d}} \le$ 60 cm cut.
    This is for the complementary method to estimate the accidental background rate.}
    \label{fig:Poisson_acci} 
\end{figure}

The spill shift method can also be utilized to estimate the rate of the accidental backgrounds
independently from the single rates method because each $n$-th spill shift case can find number of accidental background events in this control sample, individually.  
Figure~\ref{fig:Poisson_acci} shows the number of accidentally paired background events
in each spill shift cases (1,2, ... 10,000 spills shift) during 2,460,399 spills
after the $\Delta \mathrm{VTX}_{\mathrm{p-d}}$ cut, i.e.; one certain $n$-th spill shift case provides one entry in this histogram. The mean of the Poisson 
distribution is $0.494 \pm 0.008$ events over 2,460,399 spills, thus the estimated accidental background rate is $(9.24 \pm 0.15) \times 10^{-8}$ / spill / 0.75MW. This is consistent with the single rates method.

The expected oscillation IBD signal rate is 4.59$\times 10^{-8}$ / spill / MW ~\cite{cite:TDR} 
with the LSND best fit oscillation parameters. 
Thus, \JSNS  has a comparable accidental background rate as the expected neutrino oscillation signal with the LSND best bit oscillation parameters.

\section{Summary}
\label{summary}
\JSNS aims to perform a direct test of the positive result of the LSND experiment using a decay-at-rest neutrino source at the MLF and a Gd-LS detector. 
We started data taking in 2021.

The calibration runs with the accelerator scheduled timing to study the accidental background
have been performed. As a result, the single rates of the IBD prompt and delayed candidates 
at 0.75 MW of averaged beam power are 
(2.20$\pm$0.09) $\times 10^{-4}$ / spill and (1.80$\pm$0.01) $\times 10^{-2}$ / spill, respectively.
The expected rate of the accidental background is  (9.29$\pm 0.39) \times 10^{-8}$ / spill, 
which is a similar level with that of the IBD neutrino oscillation signal
with the LSND best fit oscillation parameters.
We anticipate the future improvements to reduce the accidental background with sophisticated likelihood and/or machine-learning techniques.

\begin{acknowledgements}
\sloppy
We thank the J-PARC staff for their support. 
We acknowledge the support of the Ministry of Education, Culture, Sports, Science, and Technology (MEXT) 
and the JSPS grants-in-aid: 16H06344, 16H03967 and 20H05624, Japan.
This work is also supported by the National Research Foundation of Korea (NRF): 2016R1A5A1004684, 2017K1A3A7A09015973, 2017K1A3A7A09016426, 2019R1A2C3004955, 2016R1D1A3B02010606, 2017R1A2B4011200, 2018R1D1A1B07050425, 2020K1A3A7A09080133, 2020K1A3A7A09080114, 2020R1I1A3066835, 2021R1A2C1013661 and 2022R1A5A1030700. 
Our work has also been supported by a fund from the BK21 of the NRF.
The University of Michigan gratefully acknowledges the support of the Heising-Simons Foundation. 
This work conducted at Brookhaven National Laboratory was supported by the U.S. Department of Energy under Contract DE-AC02- 98CH10886. 
The work of the University of Sussex is supported by the Royal Society grant no. IESnR3n170385.
We also thank the Daya Bay Collaboration for providing the Gd-LS, the RENO collaboration for providing the
LS and PMTs, CIEMAT for providing the splitters, Drexel University for providing the FEE circuits and Tokyo Inst. Tech for providing FADC boards.
\end{acknowledgements}



\end{document}